# Why informatics and general science need a conjoint basic definition of information

Wolfgang Orthuber

*Kiel University, UKSH*

Arnold-Heller-Straße 3, House 26, D-24105 Kiel, Germany

**ABSTRACT**
First the basic definition of information as a selection from a set of possibilities resp. domain is recalled. This also applies to digital information. The bits of digital information are parts of number sequences which represent a selection from a set of possibilities resp. domain. For faultless conversation sender and receiver of information must have the same definition of the domain (e.g. of language vocabulary). The more the definitions deviate and the greater the combinatorial freedom (e.g. at usage of vocabulary) on the way to meaning, the greater is the probability of misunderstandings. This is a serious problem not only in informatics but in all kinds of information transfer resp. conversation. Up to now the definition of the domain and of its elements is derived from context and knowledge. The internet provides an additional important possibility: A link to a conjoint uniform definition of the domain at unique location on the internet. The associated basic information structure is called "Domain Vector" (DV) and has the structure "UL (of the domain definition) plus sequence of numbers". The "UL" is not only "Uniform Locator" of the domain definition. It also identifies a certain kind of information for later comparison and search. It can be a Uniform Resource Locator (URL) or an abbreviated equivalent, e.g. a hierarchic numeric pointer or a short local pointer to a table with global internet pointers. The domain definitions can be downloaded. The DV structure can be used as general carrier of information which is language independent and more precise than language. A domain which contains DVs is called "Domain Space" (DS) and is defined as metric space. This allows similarity search according to user defined criteria, so that any kind of definable information can be made comparable and searchable in high resolution according to user selected (relevant) and objectifiable (globally uniform) criteria. DS definitions can be reused in new DS definitions. Their elements, the DVs, are automatically globally uniformly identified and defined. Obviously such conjoint definition of comparable information has great potential. It also can avoid interoperability problems and redundant programming and so save high costs.
**KEYWORDS**
Information, Domain Space, DS, Domain Vector, DV, Standard.

## 1 Introduction

Over years the global definition of information has been ignored, so this article contains critical parts. Central topic of informatics is (digital) information. There is a lot of literature about the term "information" and it is also possible to find descriptions which bring us nearer to the basis: For example "information is the answer to a question" or "information can be transmitted via a sequence of signals". These approaches already imply a set of possibilities. Sets are basal concepts. In mathematics the set concept has been proven to be essential. Here we can begin. Obviously an exact definition of information is necessary for science. We start with 2 questions:
   a)   Where is a clear and representative definition of information?
   b)   Is there (still) enough awareness of this definition?
The answer to a) can be given by looking into Shannon's well known mathematical theory of communication [18]. On the first page he wrote:

>   "The significant aspect is that the actual message is one selected from a set of possible messages."      (1)

From this we can derive the definition of information:

> **Information means selection from a set of possibilities (domain).**              (2)

The "<u>set of possibilities</u>" is called "<u>domain</u>". For faultless conversation sender and receiver of information must have the same definition of the domain. Looking at (2) we can regard information as a mapping from the domain to a subset of the domain.
So concerning a) we can conclude that the answer (2) has been given already in [18]. But concerning b) the answer is not so convenient. If there would be still enough awareness of (2), then in the internet age it would be obvious to define the "set of possible messages" resp. "domain" globally uniformly (e.g. identified and located by a global pointer, e.g. the URL of definition). This far reaching possibility is unused up to now, despite the fact that there are already publications about online defined number sequences (which represent a selection from a globally defined domain) resp. vectors (elements of metric spaces) as general carriers of information. Digital data are numbers (briefly speaking). The word "numbers" is just as general as "digital", but more meaningful, because it shows better the structure and order of digital data. The proposed data are globally uniformly defined (i.e. with the same domain). The definition is possible by all users (to get maximal competence) and the data are identified. But the proposals [9][10][11][12][13][14][15][16] have been so often and long ignored, that there must be a

*systematic cause*[1]. So it is necessary to recall (2) and to recall that a missing global (conjoint) definition of the domain causes many unnecessary and also expensive difficulties, e.g. redundant programming, interoperability problems, data silos - *missing* information exchange.

This already indicates the up to now unused potential. A description of details follows now. Concerning the comprehensive literature about "information" we restrict to representative examples. So [3][21] describe important basic concepts about information theory. Random variables with probability mass functions are used quickly at the beginning. These concepts are already introduced in former literature. So Kolmogorov wrote in the beginning of his combinatorial approach [6]:

> 'Assume that a variable x is capable of taking values in a finite set X
> containing N elements. We say that the "entropy" of the variable x is
> $H(x) = \log_2 N$
> By giving x a definite value
> $x = a$
> we "remove" this entropy and communicate "information"
> $I = \log_2 N.$'
> (3)

In other words than Shannon in [18] the term "giving x a definite value" also means "selection of x within the set X". Here we use the word "**domain**" for the (value or definition) set X, and x can be seen as random variable [3][21]. The probability mass function of random variables needs secondary statistics e.g. from multiple experiments.

So already in early literature "information" has been seen as a "selection from a domain". This definition has been used more or less indirectly also in later scientific literature. Derived concepts and phrases like "random variable" and from these derived concepts and words are common. For abbreviation purposes this can make sense *as long as* the original meaning is clear.[2] Here again we concentrate to the definition of information, especially via global predefinition of the domain (2). This is also important for design of scientific studies and for comparison and combination of results (7). The following chapters provide first hints how to proceed from the up to now usual domains of information (letters resp. characters and language vocabulary) to optimized topic specific language independent searchable domains (Chapter 3.6) of information.

## 2 UL resp. Uniform Locator is a URL or an abbreviated equivalent

We propose a data structure (5) which contains a unique pointer to a unique location on the internet. The term "URL" (Uniform Resource Locator) is widely used for this. Up to now [13][16] we also used this term, mentioning that abbreviation is possible. From now on we use the term "Uniform Locator" or "UL" for clarity. The UL can be a Uniform Resource Locator (URL) or (for abbreviation) another unique pointer within the internet. It can be also a unique number, or it can be something with the same functionality (in (5)), e.g. a byte with meaning "same UL as before" or "same concatenation of ULs as before" or a short local pointer to a local table with global internet pointers, i.e. at last to any addressable location on the internet. Important is efficiency to achieve the function: **The UL is an unambiguous and efficient pointer to the definition and simultaneously an identifier of subsequent data (numbers)**.

## 3 Domains for digital information transfer: From simple to complex

Predefinition of a domain is necessary *before* transfer of information (2) for all participants of a communication. If we want to predefine a domain, an estimation of the necessary count of explicit definitions is important for practicability. The larger the domain size (cardinality), the higher the resolution, but often also more definitions are necessary. In the worst case (e.g. not ordered language vocabulary) the necessary count of definitions is as large as the cardinality, in the best case (an ordered set, e.g. a dimension of a vector space) one definition is enough to define a domain with very large cardinality.

---

[1] Initially we assumed that science by definition is self-initiated enough to detect and multiply this. This was not the case. There have been hierarchical structures. Clear arguments up to now have not been answered by the deciding persons. In general we found very much motivation for publishing in science, but not necessarily enough motivation to follow relevance. Relevant approaches frequently need cross-disciplinary thinking.

[2] Sometimes, however, the original meaning can become hidden if derived sophisticated concepts dominate literature. This can become a problem. Usage of sophisticated concepts without clear connection to the original motivating basis can lead to large (frequently cited but) baseless constructs in science. The reader may decide, which literature about information transfer needs to recall the precondition for information transfer: A common definition of the domain (2). If possible, we avoid a more critical review of existing literature.

There is a hierarchical complexity of domains - from bits and numbers to complex meaning. Using binary system, sequences of bits can be combined to form numbers. Sequences of numbers can be combined to form vectors. If the basis of a vector space has n dimensions, the space is called n-dimensional. Every element of it is a n-dimensional vector, which typically is represented by n numbers. Numeric representation does not mean loss of generality. Even if we restrict to integer numbers within a certain range, we can use a conversion function so that every allowed integer number resp. dimension represents a well defined thing (e.g. a character or symbol if we use Unicode as conversion function). Important for expressiveness is the domain size (which is the total count of possibilities). If the domain size of every dimension is m, then the domain size of an n-dimensional vector is $m^n$. So by combining dimensions we quickly get a very large maximal domain size. Therefore the logarithmic measure (3) became standard for measuring the quantity of information [6][18].

## *3.1 Character sets as domain*

The character set is the typical, well known domain. It is used for all text of language. The ASCII code with only 128 characters shows the main advantage of this domain: Small size and small count of necessary definitions. At introduction of computers this was essential. Preknowledge (storage) of only a small set of definitions is sufficient for transfer of every text of a certain language (e.g. English). Therefore this domain is widely used. Later larger code sets like Unicode have been introduced for representation of characters from all well known languages.

Relevant disadvantage of character sets as domains is the necessity to combine many elements (characters) to get meaning. The meaning of these combinations is completely language dependent. There are many ambiguous possibilities for character combinations (from the same and from different languages) to get the same or similar meaning, for example "speak", "talk", "argue", "sprechen", "reden", "argumentieren", "说话", "تحدث", "говорить". The associated combinatorial problem is relevant. For example, if we search for a certain combination of characters, we will not find all combinations with this meaning.

A further disadvantage is the missing relation to natural order (of meaning). It makes no sense to connect "similar" characters with "similar" meaning.

## *3.2 From characters to words and phrases - language dependent step*

In case of language vocabulary the maximal domain size of the possible vectors (combinations of characters) is much larger than the domain size of the meaningful combinations (words and phrases of language). If we only allow for example integers 1..26 for representation of lowercase alphabet of the English language and only assume 5 as dimensionality (as mean word size), we get $26^5 > 10^7$ possible vectors. This is (of course) much larger than the size of a usual language vocabulary [20]. Words and phrases are the typical content of dictionaries. The combination of characters to words and phrases is the typical language dependent part in the conversion of digital information to meaning.

The domain with words and phrases of usual language vocabularies is small enough for current hardware to be predefined and represented online.

## *3.3 Words and phrases of international vocabulary as domain DL*

Words and phrases are the elements of language vocabulary. Online definition of this domain (see 3.2) means, that we could place a representative definition of the set resp. **D**omain of **L**anguage vocabulary (DL) online on a well defined location. Every element of DL would be a word or phrase e.g. in an internationally usual language like English (as reference for practicability) with additional translations in other relevant languages (multilingual definition). The simplest possibility would be a one-dimensional DL, ordered along frequency of usage, most frequent first. Every element (word resp. phrase) could be addressed by a self-extending[3] integer number l, which is the shorter, the smaller the number (the more frequent the word resp. phrase) is. Depending on l an optional second integer number m could address grammatical variations and variations in meaning. On the internet the element could be represented by the binary data structure:

    **UL** (of online definition of DL, plus) **l** (plus) **m**                                                           (4)

A sequence of (4) could be used for representation of text.
**Disadvantages** of the domain DL:

---

[3] For example an integer number i < 32 * 256^(n-1) coded from n bytes, where 3 bits of the first byte code n.

1. Special attractive software is necessary for representation the elements. The "partition of meaning" is different among different languages, so we cannot expect to find a direct translation of every word. Because the meaning depends on context, it is more realistic to expect a translation of words and phrases to words and phrases like in dictionaries. This increases the count of elements in DL. So DL has larger cardinality and is more difficult to handle, compared to the in chapter 3.2 described domain "characters". For practicability DL has to be downloaded at least in English and system language.
2. Definition of DL is elaborate and no initial task. The benefits of DL depend on its richness of detail and scope. This can increase step by step. The more work is put into the definition of DL, the clearer become its advantages.

**Advantages** of the domain DL:
1. More language independence: Because the elements of DL are defined after the language dependent step 3.2 , we get more language independence. During composition of a text (stringing together words and phrases) in the own (system) language the author automatically can get information e.g. about the English variant of the words and modify m (see (4)) so that the variant is closest to the intended meaning to get "international" meaning.
2. Clarity and precision: Redundancy can be avoided from the beginning. Step by step the domain DL can be refined, e.g. by phrases which represent more complex special cases, so that this works better and better and also provides more clarity and precision than conventional language vocabulary.
3. Definition of similarity: It is possible to connect and group together synonyms within the same element of DL and distinguish these only by variants of m, see (4). Also elements with similar meaning, for example the words "go" and "run" and "drive" can be grouped together. For quantification of similarity further dimensions can be added. At this it is also quantitative parameters like "speed in meter per second" can be used. Similarity comparison and search of such ordered parameters is possible using a metric (distance function, see [23]), e.g. Manhattan metric. After this it is possible e.g. to search all words "drive" with speed (in meter per second) between 100 and 120.
4. Reusable: The better the elements of DL are defined (step by step) e.g. concerning clarity and precision, the more attractive these are for precise texts and also for reusage within more complex (multidimensional) definitions.

## 3.4 Conversion of characters to international vocabulary DL

We are accustomed to combine characters to more complex units. Sometimes the connected freedom is intentionally used to build new words, e.g. new names, so that these can be recognized later again. So it is necessary to include this option. For this initial free input of characters in system language can be designed like today. After input of text the software can optionally (after confirmation) replace recognized words and phrases of system language by equivalent elements of an international vocabulary (3.2). In case of ambivalence the software can offer several choices. So shortly after input the writer can design the own text more in internationally unambiguous way. It depends on the purpose of the text and the intention of the writer, how far is gone towards such precision.

## 3.5 Words and phrases combined

If w is the cardinality of the domain DL with language vocabulary and if it is combined n times, so the cardinality of the combination is $w^n$. This makes at once clear, that before trying to predefine online a combination with n≥2 (without consideration of meaning) it is much more efficient to concentrate on n=1, i.e. on predefinition of all uncombined multilingual language vocabulary and on its extension with additional meaningful phrases in one domain.

More complex meaning we can predefine *explicitly* online at addressable location. In the World Wide Web all meaningful (and other) content can be represented by a pointer or *link* to its location - using the well known URL [1] [2] or the UL (see chapter 2). This is also an example of (2): The link represents a selection from a domain which is the set of all possible links. Every link represents certain addressable content. This usage has proven to be very successful.

## 3.6 Domain Spaces (DSs) as domains with Domain Vectors (DVs) as elements

A link or UL can also locate the definition of another domain - it can be seen as a "*link to a space*". After this usually further selection (within the defined space resp. domain) is necessary. This is not trivial, but this also offers completely new opportunities. To enable the additional possibility of user defined similarity search in the domain, it is advantageous to define the domain online as metric vector space, see (6) and [23]. We call it "**Domain Space**" (**DS**). Every element of a DS is called "**Domain Vector**" (**DV**) and has two components:
1. The UL of the DS definition, which according to chapter 2
    a. identifies the data (by identifying the containing DS) and which
    b. locates the definition of the data (by locating the definition of the containing DS),
2. and a sequence of numbers, which represents the selection within the DS.

From this follows the structure of the DV:

**UL** (of online definition, plus) sequence of **numbers** (5)

This binary data structure is a generalization of (4) and also consequently follows the definition (1) of information. It is a selection from a set of all DVs. For this first (see 1) the DS is selected via (a sequence of numbers which represents the) UL of its definition, and then (see 2) via sequence of numbers the DV is selected within the DS.

### 3.6.1 Disadvantages of the DV data structure
Also today all digital information consists of sequences of numbers.
1. Compared to the up to now usual structure "sequence of numbers" of digital information, the main disadvantage of the DV (5) is the additional space requirement by the UL, which can be minimized, see chapter 2.
2. The DV (5) is dependent on online definitions. Therefore download of used parts of online definitions is necessary. There should be an infrastructure which guarantees stability of online definitions, e.g. by mirroring.

It may be also argued that work is necessary for making the online definitions. But instead of this today we make definitions by context in uncontrolled varying way again and again together with the data. This also makes later comparison of data more difficult or even impossible.

### 3.6.2 Advantages of the DV data structure
For the space needed by the UL we get a global pointer to the online definition (which usually is much larger than the UL) of the following data (numbers). The definition can be also multilingual. Simultaneously we get a global identifier of the data which makes these comparable by machines. The advantages of this can be listened here only partially, see also [13][16]. The Domain Vector (DV) structure (5) is advantageous due to many reasons:
1. The data structure (5) enables the combination of maximal competence (of all users which can define via UL) with maximal efficiency (number sequence allows free definition of every bit).
2. Every DS with *all* its addressable elements resp. DVs has an online definition at well defined location. If necessary, the definition can be extended afterwards. DVs can be used as pointer to explicitly defined DS elements (e.g. a multidimensional table or ontology - like DL in chapter 3.3).
3. As "Domain Space" or "DS" the domain is a metric space. This means that a metric resp. distance function F between DVs is defined. Let DV1, DV2, DV3 be any elements of DS. Then as metric F fulfils [23]:
    $F(DV1,DV2) = 0$ if and only if DV1=DV2
    $F(DV1,DV2) + F(DV2,DV3) \geq F(DV1,DV3)$ (triangle inequality)
    $F(DV1,DV2) = F(DV2,DV1)$ (6)
   There are useful and uncomplicated functions which fulfill this, e.g. "Manhattan distance" or "Euclidean geometry". The function F can be adapted to the requirements. For example it is recommendable that most important criteria (dimensions) mostly contribute to the value of F. Later F can be used for similarity search: The smaller F(DV1,DV2), the more "similar" are DV1 and DV2. So via this metric the users (owners of ULs) can define search within the own DS. All kind of quantitative data can be directly defined as dimensions of a DS and so are available to similarity search [23].
4. Existing definitions can be reused within new definitions (nesting of DS definitions) [16], so that complex search over multiple DSs becomes possible.
   If, for example, there is a one dimensional DS with dimension "length in meter" and another DS with dimension "width in meter", these definitions can be included in more complex multidimensional DSs definitions. Then it is possible to search within *all* these DSs for all DVs with certain length and width.
   The possibility of reuse also shows that usage of all online DS definitions should be completely free. The term "DS definition" implies free usability, because restrictions by legal constructions would open a way into unnecessary complications.
5. The definition of a DS is adaptable to the needs - from simple to complex. Additional dimensions can be appended afterwards.
6. If the value set of a DS dimension is known exactly a priori, a very efficient definition is possible. For example in case of only 2 alternatives (yes/no) we need only 1 bit. If (the range of) the value set of a dimension is not known a priori, the first byte of a number (mantissa and exponent if necessary) can also contain bits with length information. Such a number can adapt its length to the requirements (self-extending number[3]), to minimize the count of unused bits.

7. Generation of DS definitions can be automated e.g. by integration into programming languages.
8. The identification of DV data by UL makes these interoperable, globally searchable and comparable. This is generally important in informatics and for objectifiable information exchange, e.g. in science, see 5.

Further features of DSs as domains are described in [13][16], also similarity search and the search engine http://numericsearch.com, which introduces the principle using a local database, see Figure 1, Figure 2, Figure 3.

**Figure 1: Online search engine within local database for demonstration purposes. Domain Spaces (DSs) can be defined and filled with data by users. Every DS can be selected by clicking on its index i7.**

**Figure 2: DS 1008 has been selected. Then data for exemplary similarity search in 2 dimensions has been provided.**

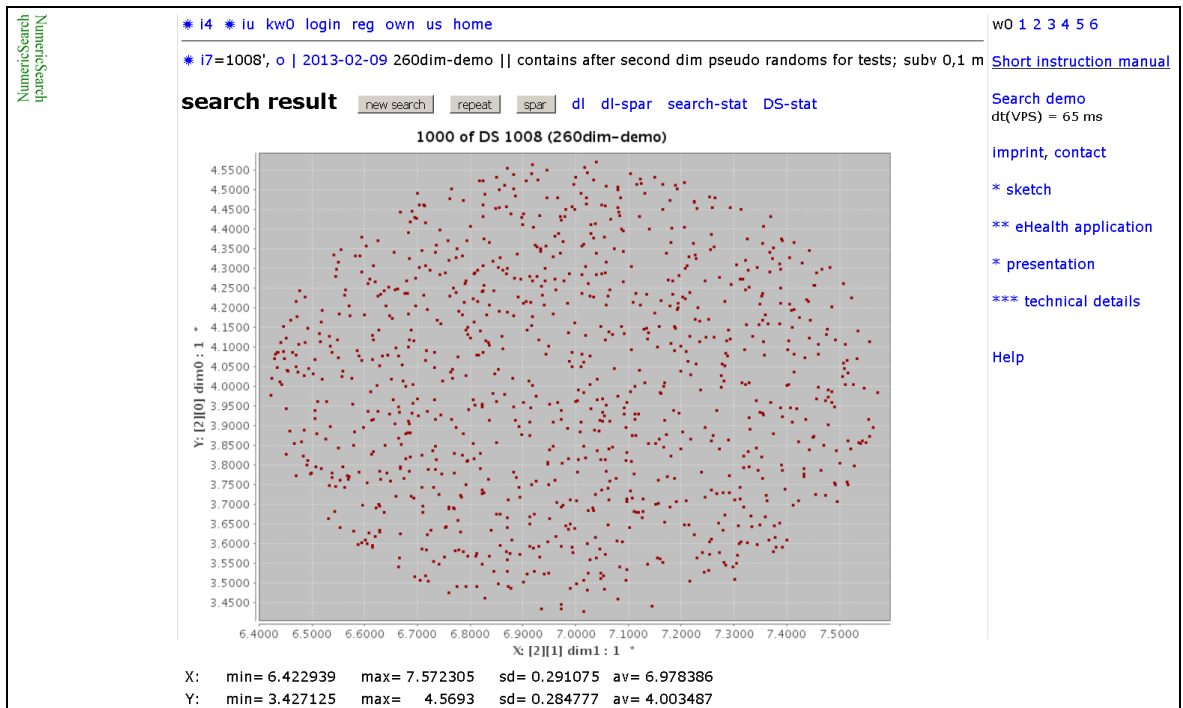

**Figure 3: Graphical search result after input of Figure 2. Every found DV is represented by a point. Its coordinates show the values of the searched dimensions in this DV. The DVs of DS 1008 contain equally distributed pseudo random numbers 0..10 in these dimensions and Euclidean distance was chosen as distance function. Because in the graphical search result the 1000 (DVs with the) most similar points are shown, i.e. those with smallest distance to the (DVs with the) searched point (7,4), an elliptic distribution around this point results.**

### 3.6.3 How to get high resolution and also enough range of DSs

DS definitions can be high dimensional and can describe complex situations. To achieve high resolution and simultaneously enough range in meaning, a lot work has to be put in the development of online definitions. This work can be done by all users due to usage of a global pointer (UL in (5)) in every DV.

There are 2 principle means to get besides high resolution also enough range in meaning:
I. Predefinition of many dimensions in a DS: This definition quickly leads to large cardinality (see chapter 3) and so allows combining high resolution with high range. Enough preknowledge about possibly interesting dimensions is necessary. It is recommendable to include existing definitions via nesting (see 4).
II. Belated combination of DVs from different DSs: Combinations of DVs is always possible like combinations of words in text. The advantage is more flexibility, the disadvantage is less unambiguity: It is possible that different combinations lead to the same meaning. But DVs can be defined more complex than simple words, so that so that already small combinations of these lead to enough meaning.

In the initial stage, meaning is typically achieved by combinations of DVs. Later more complex and high dimensional DV definitions are available, so that the part of belated DV combinations can be reduced more and more.

### 4 Definition of specialized Domain Spaces (DSs)

The DV data structure (5) does not restrict freedom of application. For example the first number after the UL can contain the count of the following numbers. If e.g. these numbers are defined to contain Unicode values, the DV (5) can contain free text. But in this case we still have the language dependent step 3.2 and a lot of combinatorial freedom before we get meaning. Usage of an international vocabulary (see chapter 3.3) reduces unnecessary combinatorial freedom, but much more can be done by topic specific DSs:

Whenever possible we recommend specialized online definitions which avoid unnecessary combinatorial freedom when the DV data structure (5) is filled with data (numbers). For this it is advisable to predefine for specific topics and domains as much as possible meaning carrying variables as dimensions of combinable specialized DSs, so that "similar" meaning leads to "similar" results (5). Later such results are directly comparable and systematically searchable. Obviously this is important for scientific, technical and other professional literature. For such special communication which is focused on a certain topic or domain, according to chapter 3.6 predefinition of specially designed DSs can save a lot of detours caused otherwise by language and/or incommensurable standards. These can be used directly (I) or combined (II).

When defining a DS for a certain topic or domain, it is desirable to cover as much as possible breadth in meaning by *independent* variables resp. dimensions. Later this facilitates search. Often many details are important in a certain subject matter and it is useful to define appropriate dimensions for description of such details. It is advantageous to reuse already existing definitions via nesting. This can lead to definition of high dimensional DSs. But this does not mean that all these dimensions are used in every case. The dimensions are by default only optional container for information. The user, who selects a certain DS for information transfer, can decide which dimensions are filled with values and which not, according to the requirements. The better the DS fits to the requirements, the more information its dimensions can transport in efficient, reproducible, comparable and searchable way. In the course of time more and more such predefinitions can be made available which allow increasing the directly comparable part of exchanged information.

## *4.1 DSs for decision support*

Global definition and precision of DSs is important for sharing of experiences, e.g. in medicine. So a typical nontrivial application of specialized DSs is decision support. Experts about a certain subject area can share their experiences and together develop the best definitions of their DSs. The DS contains:

i. Dimensions about preconditions of a decision. Interesting are all parameters with relevant influence on decision and its result.
ii. Dimensions about the decision. Interesting are all relevant parameters which describe a possible decision.
iii. Dimensions about the decision result. Interesting are all parameters which describe relevant consequences of a decision.

Usage: Before decision most interesting parameters resp. dimensions for descriptions of the current situation (preconditions) i are selected and together with varying decision parameters ii provided to the search engine, which returns for every decision variant a group of cases with "similar" dimensions (Figure 3) and for every group means and standard deviation of the result dimensions iii. The better the dimensions i and ii describe the situation at time of decision and the more similar the found dimensions are and the larger the found group is, the more reliable the means of the result dimensions iii represent the expected outcome of a decision variant. The statistics recommends the decision variant with the most favorable result dimensions iii.

## *4.2 DSs for searchable description of selectable alternatives*

Of course things are more trivial than in 4.1, if relevant decision results directly follow from a decision. For example if the decision is simply purchase of an article, its relevant technical data can adequately describe a relevant (part of the) decision result iii. We can simply search for a desired decision result to find the desired decision. Relevant preconditions can be included in a multidimensional search, if wished. These can be also belated combinations II of DVs.

## *4.3 Content of DS definitions*

The destination address of the UL in the DV (5) points to the entry of a binary online definition of a DS. Here at first a self-extending[3] version number is appropriate, and then further content in dependence of the version number, e.g. a list of pointers to multilingual definitions of every dimension (number), further pointers to graphics, subprograms for handling the data with optional source codes etc.

Trying to list the possible content of a DS definition would exceed the scope of this article. A small excerpt of possible content shows Figure 4 with an input window of our prototype for definition of a dimension of a DS, [13] contains some additional information.

**Figure 4: Exemplary definition of the dimension "Price" in the Domain Space "Cupboard" (DS 1006 in Figure 1).**

## 4.4 DS based search compared to word based search

Already a single DS can contain a complete language vocabulary and is free for extension. DL in chapter 3.2 is an example for such a DS. Also (4) is free for extension - one element represents not only one point but can be extended to a multidimensional metric space. The fact that in example (4) simply more dimensions can be added provides first insight in the the much higher cardinality and resolution of DSs than language vocabulary. This principle can be extended to all user defined topics. The users know best about their special field. They can first look on the internet for existing (reliable, mirrored) definitions concerning it and can use these completely for own DVs, or reuse these in new DS definitions together with new own definitions. The users can best define dimensions of DSs which describe the most interesting criteria in their special field. DSs are definable by users, so also DS based search is definable by users.

If, however, a new online definition is not wished, then it is also possible to combine explicitly DVs as described in chapter 3.6.3, II. Later it is also possible to search for such explicit combinations. Word based search cannot nearly achieve resolution and precision of DS based search.

The technical possibilities are so far reaching, that it should not be expected that every advanced search can be done free of charge. Transparent costs for searching data could be also seen as step towards more honest distribution of expenses.

## 4.5 Example: Domain Vectors (DVs) in medicine

An important application of the above decision support scheme 4.1 is medicine [15]. DSs about medical findings, about treatment (kind, count, frequency, dose etc.) and treatment results describe preconditions i, decision ii and result iii. Of course it is relevant for medicine, that the definition of the DSs resp. data (5) (DVs) is inherently international: This makes medical experiences internationally comparable. There are a lot of decision relevant data in medicine, also e.g. feature extraction results of images and other complex findings. Original data are quantitative. So a general approach for making quantitative data searchable is necessary, because searchable data are essential for decision support. This was one of the first motivations for development of this concept[4]. Machine readable and (via UL) addressable online definitions are a general practical approach. Existing nomenclatures like LOINC [7] are convertible in such uniformly identified online definitions. Of course these should be completely free (see chapter 3.6.2, 4).

Like described in [15], the abbreviated procedure for decision support in medicine is:

a) The physician provides an initial diagnosis (e.g. from ICD-10 [22]) to the decision support system, which answers by showing the most frequent (dimensions of DSs which represent) further diagnostics resp. measurements (inclusive data about treatment) made by colleagues under this condition.
b) The physician decides about further diagnostics. Because these have unique names and because the system is connected, after completion of further diagnostics and measurements all results can be provided automatically to the system.
c) The physician decides about most interesting results $x_k$. The system shows to every $x_k$ the standard deviation $s_k$ and suggests intervals $I_k := [(x_k - r_k s_k), (x_k + r_k s_k)]$ for search. At this the physician can modify the $r_k$ and also shift some $I_k$ e.g. for testing variable treatments. After this the intervals $I_k$ together describe a search command which is sent to the system.
d) The system searches the group G with all patients whose $x_k$ lie within the $I_k$ and can make further statistics within this group. The system shows among others to chosen diagnostic and dependent parameters $x_l$ the means $m_l$ and standard deviations $d_l$ . These can be also descriptions of treatment results. So by modifying the $I_k$ in c) the effect of different treatments on the treatment result can be checked within the individual group G "near" to the individual patient.
It is also possible to make similarity search as shown in Figure 3. For similarity search e.g. the inverse widths of the intervals $I_k$ in c) can be used as weights of the dimensions in the distance function f, as condition for "enough similar" e.g. a maximal Euclidean distance can be given. Figure 3 illustrates a representative result for such search in case of 2 dimensions.
e) Using the provided information and optionally further results it is possible to continue at b) or even at a), to find the treatment with the best results.

### 4.5.1 Distribution of search and anonymization of search results

The search d) can be distributed. It is not necessary to collect all patient data within a central database. Important is, that the definition of the data is identical, i.e. that there is one and the same definition of all data (of every used dimension). This is automatically guaranteed by the DV data structure (5) . Under this condition it is not even necessary to use the same search engine. The search parameters from c) can be distributed worldwide to several databases with own search engines. Every search engine responds to the sender of the search parameters c) back with the anonymized statistics d) over the found group G as search result within the own database. All these are collected at the sender of the search request c). There every search result is weighted by the size of the found group for calculation of the combined statistics over all found patients. So we can get a worldwide anonymous statistics over all patients. So the data of the patients need not be collected in a central database, it is sufficient to provide the data of every patient to the same local database d) which can answer anonymously to world wide requests c).

---

[4] The basic ideas have been published already many years ago [9][10][12][13][15][16]. These have been ignored up to now, the connected technical potential for improvement of information exchange in medicine has been delayed for years. Obviously this is not without consequences. The example "medicine" shows that there are relevant reasons to become more critical (see chapter 1).

## 4.6 Example "DV money": Domain Vectors in a digital money system

Concerning digital money there is already a lot of literature about security and cryptography. This is *not* focused here. Focus of this chapter is design of digital money for efficient control.

There are very controversial opinions about design of money. Hence it is impossible to make a proposal which is favored by all. But there are objectifiable arguments as foundation of the following proposal, which is described here in abbreviated form.

### 4.6.1 Why in future efficiently controllable digital money is important

Today most money is not generated by the central bank but as bank deposits from commercial banks through borrowing and lending. From this a series of investments and reinvestments is generated, where the influential investors are (influential because they are) interested in as high as possible individual return of money on investment of money. This would be a good idea, if maximal individual return on investment of money would be *coupled with reality*, i.e. if it leads to valuable (long-term) future. But precondition for a valuable future are also investments in projects concerning tasks which cannot provide (quickly measurable) individual return on investment, e.g. investment into education, infrastructure (e.g. of healthcare), waste prevention [19], generally conservation of the environment, general reserves for pension, development aid (also against population explosion), and so on. It is already noticeable that today these tasks tend to be neglected, and that this leads into a deadlock. Alone because of population explosion [5] indirectly the probability of conflicts (ruined future) is increasing to a dangerous level.

Therefore we can summarize: **The large scale experiment with not enough controlled money already has been done and the result is underfinancing of essential global tasks without quick individual return on investment.**

Therefore more or less control is necessary. This is typically done by a government using tax ("Steuer") to control ("steuern") the market. **Here the question is how to create best basic technical preconditions that government gets information and then can define tax so that investment into (long-term) future can be optimized, with minimal bureaucracy, to the best of the available information.** So the first step is not intervention. First it is necessary to get well founded (objectifiable) information about the market. If intervention is necessary, it should be well founded. From this the aim of the proposal is derived:

We want to provide a (initial, summary) technological description of "DV money": a **digital money system which can be controlled automatically and efficiently with minimal bureaucracy**. For privacy the system by default protects personal information. It provides anonymized statistical information according to definition of the government. The focus and resolution of this information fully depends on the intention of the government which defines the collected statistical data. All definitions are published online and so are usable globally. This is important for objectivity and for international harmonization of tax. The system also provides efficient means for automatic tax estimation and for definitive tax calculation and collection by tax authorities.

### 4.6.2 DV money: Digital money system with DVs

As mentioned above, the proposal starts with current preconditions. First the infrastructure is adapted and information is collected. Only after getting enough information new tax is introduced and old tax removed stepwise and carefully. Loopholes can lead to dysfunctions. Therefore the DV money system should be broadly accepted. Settling of differences will take more or less time. Nevertheless there is a need for a commonly controlled money system and it is important to know a concrete efficient technical possibility, to start discussing how to go in this direction. Within the scope of this article we provide a summary of the technical and organizational details (in paragraphs):

1. A trusted and secure digital environment is introduced. All accounts are licensed within this environment. Cash is exchanged into deposits on the accounts. Then cash is abolished and payment is done in the secure digital environment. Due to the expected high volume of traffic: Low energy costs are an important criterion already during design, e.g. at selection of algorithms. Efficient strong encryption is recommended. It is, however, not necessary and very inefficient (and therefore no long-term solution) to use for example a blockchain technique like in the digital cryptocurrency Bitcoin [8].
2. According to the needs the online definitions of the (tax DSs which contain the) tax DVs are (if possible, by usage of already existing international definitions) provided by the government and internationally usable.
3. It is comfortable to provide every tax payer with enough accounts - at least one account for every type of income resp. tax differentiation. Every such account is labeled by a DV (5) which is called "tax DV". Due to its concentrated information less than 50 bytes suffice for a basic tax DV. Also a more comprehensive and nested tax DV is possible, depending on the tasks. The tax DV could contain for example a pointer within expandable international tax ontology, GPS coordinates etc.. It is most efficient to code also all other account and bank transfer information as DV, e.g. date, balance, amount of every transfer, IBAN etc.
4. Transactions are labeled by tax DVs. These provide enough information for automatic tax calculation (together with already given information) and are provided on the invoice or by the system. Transactions to accounts with tax DVs by default inherit this tax DV (automatic labeling). Because every taxpayer has an account with specific tax DV for every type of income, an automatic reality conform differentiation and taxation of transferred money according to definition of tax DVs is possible.
5. It is usually sufficient that every tax payer is registered pseudonymously in the DV money system, and labeled by a DV or DV group. The pseudonymous information is used by the tax authorities and allows unbureaucratically quick estimation of tax. For automatic tax calculation the information of all DVs (of tax payer, transactions, and accounts) is used.
6. For governmental control of the market only anonymous information is necessary. It is sufficient that the system provides (anonymous) statistics over selectable groups of (pseudonymous) tax payers. The criteria of these groups depend on the DV definition and the selection criteria. By law a minimum of (change of) group size of statistics can be defined for anonymization purposes.

### 4.6.3 Most relevant disadvantages and advantages of DV money

**Disadvantages:**
1. There will be objections concerning privacy, because accounts with DV money must allow controlled anonymized access by the government and pseudonymized access by the tax authorities. But also today legal money is not hidden from access by state authorities, and the DV money system provides well defined rules for this. So security and privacy must have high preference. The system must provide quick pseudonymization for tax authority and anonymization for general inquiries (by restriction to statistical output, see 4.6.2, 6).
2. For automatic tax calculation it is necessary to introduce per taxpayer at least one account with tax DV for every (by tax differentiated) type of income. Further accounts may be necessary in case of more complicated tax. In case of a company with very much different kinds of income many accounts may be necessary. But these accounts also introduce more clarity and the option for automatic tax calculation. The technical costs of every account are small, because its memory requirement is small. In case of a typical employee one account with tax DV is sufficient anyway, because there is only one type of income (wage).
3. Loopholes would lead to dysfunctions. Therefore a broad, at best international acceptance is necessary.

**Advantages:**
1. Efficient solution: As described in chapter 4.6.1, uncontrolled financial flows lead in the long run to severe problems due to neglection of global tasks which provide no quick return on investment, but which are essential for long-term future. Therefore more or less control (e.g. by national and later international government) is necessary, and an efficient solution is desirable.
   The DV money system allows very efficient control due to the natural efficiency of DVs. The tax DVs are defined according to the wishes of the government. Tax can be modified carefully under observation of the market, to keep the system predictable and attractive for future orientated investment.
2. The DV money system is widely adaptable to the needs. Initially it can be used only for collection of statistical information according to the definition by the government, later it can be also used for more or less control by tax. First the definition of tax DVs can be adapted to the existing tax system. The resolution of this must cover at least the resolution of existing tax differentiation. Then the definition of tax DVs can be refined and more adapted to the requirements. Step by step current tax can be replaced by tax from accounts with tax DVs. At this during introduction of the new money system the amount of tax can remain equal. So the DV money system can replace the existing system without change of tax, but with much more efficient and automatic tax calculation.
3. The government automatically gets more precise statistical, anonymized information about cash flows and the market than today. Also the reaction of the market to certain changes of tax can be observed much better than today, to detect possible need for correction of tax much earlier than today. So the probability of errors by the government when defining tax can be reduced due to early and precise information from the market, which can be also scientifically analyzed.
4. The effect of decisions about tax is controlled directly and quickly. Example: Suppose that the product XX causes long-term damage. Today in special cases there is an extra tax on such a product, e.g. on tobacco. But this requires additional bureaucracy. In the DV money system this product group is simply one element of many elements in the tax DS. (The element can be a group of elements, because an additional dimension of the tax DV can be used for further differentiation and definition of additional subgroups). This means, that there is a certain tax DV (group) called "tax DV_XX" which can be used as marker of this product (and of additional subgroups of) XX. Every manufacturer of XX has an account which collects income from sale of the product XX. In case of DV money this account is market by tax DV_XX. So a simple search for all (accounts with) tax DV_XX can collect all data about income of DV money from product XX. The government can decide about certain tax on all these accounts which then usually can be collected automatically due to differentiation of accounts by tax DVs. The result can be controlled directly by search of all accounts with tax DV_XX. The sum of income over these accounts sums up the total income of DV money from sale of product XX.

If the government wants to promote another useful product YY by subvention, this can be realized analogously like described above by definition of "negative tax" on accounts which are marked by the (to product YY associated) tax DV_YY.

5. The possibility to make easily interventions with DV money does not mean that interventions are done. Initially (by default) DV money makes no intervention. Intervention starts as soon as certain tax DVs are coupled with nonzero tax.
6. Like DSs and DVs, also tax DSs and tax DVs are automatically defined globally and can be defined multilingually. This facilitates international harmonization of tax. An international organization can early create an expandable multilingual international tax ontology, which later can be used by national governments, which - if wished - can contribute to expansion of the international ontology or add own extensions.
7. The adaptability of the DV money system allows to use it as a potent tool of the government for detailed analysis and - if wished - control of financial flows. Reversely also the actions of the government can be controlled by people due to systematically searchable and transparent publication of all tax DV definitions and of tax. The tax DVs can be defined very precisely (e.g. contain GPS coordinates of the tax payer). If at this certain (groups of) tax payers are preferred by arbitrary (unexplained) decisions about tax, this becomes visible. Today arbitrary (unexplained) decisions of a government can prefer certain groups in non-transparent way. The DV money system can provide an appropriate and searchable infrastructure, that decisions of the government about tax DVs and tax can be published in more systematic and transparent way, together with explanations and justifications of the decisions. So possible abuse is easier detectable.
8. Today a lot of work is invested into analysis of the market, to get (high) income by extraction of capital by quick investment, sell and reinvestment (trade), without consideration of effects on *real* long-term future. As in chapter 4.6.1 described, this results automatically in underfinancing of essential global tasks without quick individual return on investment. The DV money system can be designed to make investments with positive effect on common future more attractive than other investments.
9. In the DV money system illegal actions are easier detectable, at least as soon as illegal valuables need to be exchanged (back) into DV money.
10. The costs for cash will disappear and all payment can be done via Smartphone and/or card.
11. The new information provided by the DV money system can be used for scientific analysis, e.g. for prediction of the market in case of certain modification of tax, in dependence of context and global situation.

### 4.6.4 Summary to the DV money system

According to 3 in chapter 4.6.2, every tax payer gets at least one bank account for every type of income which is differentiated by tax. A tax DV is combined with every such account and transactions to this account automatically inherit this DV for quick tax estimation and later tax calculation. Using anonymous statistics the system can be designed "minimally invasive" so that only necessary information is extracted for efficient governmental control of DV money. Independently of this pseudonymous information can be used for unbureaucratic tax calculation.

The DV money system also provides new detailed statistical information for scientific analysis to find objectifiable approaches for careful optimization of tax DV definitions and of tax, according to the requirements, for improvement of common long-term future.

# 5 Why informatics and general science need a conjoint basic definition of information

It is well known that the costs and inefficiency caused by interoperability problems and redundant programming are very high. So the advantages of the DV data structure (5) for informatics are obvious: The UL is a global identifier of the data and simultaneously a global pointer to the machine readable definition. At this efficiency is central focus: The UL allows maximal abbreviation (chapter 2) and the treatment of data as number sequence allows definition of every bit in maximal freedom. Due to exact identification of the data by the UL the appropriate software can be selected automatically to avoid interoperability problems. The software can be made available online, if wished with source code. It can be locatable e.g. by pointer from the DS definition or by using the UL as identifier. So software for this kind of data is searchable. This is important to prevent redundant programming. Data with the same UL can be handled in distributed systems (e.g. search engines) with the same algorithms. Chapter 4.5 shows a medical application. The algorithms can be also made available online as software modules. The output of appropriate algorithms can be combined, if wished. Due to machine readability of online definitions, of data and even of associated program modules, a lot of time can be saved.

There is a large scope for development also in general science: The (on the internet simple) global definition of the domain (2) has been neglected over decades, else today in different areas of science the standardized global online definition of domains for language independent objectifiable information exchange would be usual, because

> "science" by definition deals with "objectifiable" resp. "globally comparable" information         (7)

Online definition of domains would be done in uniform way in dependence of the subject area. So important configurations could be predefined systematically (e.g. by a suitable professional organization of specialists) as comparable and searchable elements of a domain which is defined online as metric space (see 3.6). At this definition of similarity (for comparison and search) can be adapted to the needs and interests. The optimization of the different (online defined) domains would be an important topic in science[5].

The large accumulation of publications is increasing and more and more difficult to survey. Without conjoint definition of information e.g. the results of medical studies are usually not directly comparable. A time consuming Meta study can improve the situation in case of partially comparable results, but it cannot directly combine these [17]. If, however, the results have the same definition and are published online in machine readable form (5) together with statistical weight, even an automatic comparison and combination is possible.

Here we cannot list all possible benefits of conjoint definition of information. If information is carried as DV of a well defined meaningful DS (see 3.6), it can be compared by machines and therefore divergent information is easier detectable by machines. So in case of a well defined environment (4.4) it can become difficult to publish wrong facts unnoticeable. Certainly authors can refuse application of DVs (5). But as soon as these information carriers are fully established, their refusal is also a hint to the reader. The reader can decide to look only to literature, if it contains DVs of certain (well defined) DSs, i.e. DVs with certain ULs (5). If for a relevant topic there are still no DSs, these can be defined online.

---

[5] Today important "similar" configurations are described again and again in scientific literature by help of language, using more or less different formulations (see 3.2), in this or that way. As consequence the results are more or less comparable or not comparable. Automatic comparison and collection of results is not possible. There would be a much better situation if these configurations are already adapted during design of a scientific work to the available best online definitions, and if as consequence the results are also online and machine readable, automatically collectable, comparable and searchable. There could be much better preconditions (see e.g. chapter 4.5) for utilization of scientific research and literature.

# 6 Conclusion

Information means selection from a set of possibilities (domain). Predefinition of the domain (2) is necessary *before* transfer of information for all participants of a communication. The internet offers a new technical possibility: Via UL (Uniform locator, chapter 2) accessible online definition of the domain. The data structure is called "Domain Vector" (DV) (5). The definition is automatically worldwide (globally) valid and the domain can be defined as metric space or "Domain Space" (DS), so that user defined similarity search is possible in the DS. Its elements (DVs) are globally identified by the same UL in (5) and selectable according to (2) by a sequence of numbers. This leads to the proposal of the DV data structure (5) which allows the combination of maximal competence (due to definition by all users via UL) with maximal efficiency (number sequence allows definition of every bit in full freedom).

Obviously such global definition of comparable information has great potential in general and in science, of course also in informatics (chapter 5). For example it can avoid interoperability problems and redundant programming and so save a lot of unnecessary costs.

## REFERENCES

1. Berners-Lee, T., Masinter, L., & McCahill, M. (1994). Uniform resource locators (URL) (No. RFC 1738).
2. Berners-Lee, T., Dimitroyannis, D., Mallinckrodt, A. J., & McKay, S. (1994). World Wide Web. Computers in Physics, 8(3), 298-299.
3. Cover, T. M., & Thomas, J. A. (1991). Elements of information theory. John Wiley & Sons.
4. Dietze, S., Benn, N., Domingue, J., & Orthuber, W. (2009). Blending the physical and the digital through conceptual spaces.
5. Ehrlich, P. R., & Ehrlich, A. H. (1990). The population explosion. New York, Simon and Schuster, 320 p.
6. Kolmogorov, A. N. (1965). Three approaches to the quantitative definition of information. Problems of information transmission, 1(1), 1-7.
7. McDonald, C. J., Huff, S. M., Suico, J. G., Hill, G., Leavelle, D., Aller, R., ... & Williams, W. (2003). LOINC, a universal standard for identifying laboratory observations: a 5-year update. Clinical chemistry, 49(4), 624-633.
8. O'Dwyer, K. J., & Malone, D. (2014). Bitcoin mining and its energy footprint. ISSC 2014/CIICT 2014, 2014 p. 280 – 285
9. Orthuber, W., Fiedler, et al. (2008). Design of a global medical database which is searchable by human diagnostic patterns. The open medical informatics journal, 2, 21. https://www.ncbi.nlm.nih.gov/pmc/articles/PMC2666959/
10. Orthuber, W., et al. (2009). A searchable patient record database for decision support. In MIE (pp. 584-588). https://www.ncbi.nlm.nih.gov/pubmed/19745378
11. Orthuber, W., & Dietze, S. (2010). Towards Standardized Vectorial Resource Descriptors on the Web. In GI Jahrestagung (2) (pp. 453-458). http://cs.emis.de/LNI/Proceedings/Proceedings176/453.pdf
12. Orthuber, W., & Papavramidis, E. (2010). Standardized vectorial representation of medical data in patient records. Medical and Care Compunetics 6, 153-166.
13. Orthuber, W. (2014). Uniform definition of comparable and searchable information on the web. arXiv preprint https://arxiv.org/abs/1406.1065.
14. Orthuber, W. (2015). How to make quantitative data on the web searchable and interoperable part of the common vocabulary. In GI-Jahrestagung (pp. 1231-1242). http://cs.emis.de/LNI/Proceedings/Proceedings246/1231.pdf
15. Orthuber, W. (2016, September). Collection of Medical Original Data with Search Engine for Decision Support. In MIE (pp. 257-261). http://www.orthuber.com/SHTI228-0257.pdf
16. Orthuber, W., & Hasselbring, W. (2016). Proposal for a new basic information carrier on the Internet: URL plus number sequence. http://oceanrep.geomar.de/34556/1/ICWI2016.pdf
17. Paterson, B. L., & Canam, C. (2001). Meta-study of qualitative health research: A practical guide to meta-analysis and meta-synthesis (Vol. 3). Sage.
18. Shannon, C. E. (1948). A Mathematical Theory of Communication. Bell System Technical Journal. 27 (3): 379–423.
19. Singh, J., Laurenti, R., Sinha, R., & Frostell, B. (2014). Progress and challenges to the global waste management system. Waste Management & Research, 32(9), 800-812.
20. Smith, M. K. (1941). Measurement of the size of general English vocabulary through the elementary grades and high school. Genetic Psychology Monographs, 24, 311-345.
21. Wang, S., Fang, Y., & Cheng, S. Basic Information Theory. In: Distributed Source Coding: Theory and Practice (2017), 263-291.
22. World Health Organization. (2004). ICD-10: international statistical classification of diseases and related health problems: tenth revision.
23. Zezula, P., Amato, G., Dohnal, V., & Batko, M. (2006). Similarity search: the metric space approach (Vol. 32). Springer Science & Business Media.